# Shaping symmetric Airy beam through binary amplitude modulation for ultralong needle focus

Zhao-Xiang Fang,[1] Yu-Xuan Ren,[2,a)] Lei Gong,[1] Pablo Vaveliuk,[3] Yue Chen,[4] and Rong-De Lu[4,b)]

[1] *Department of Optics and Optical Engineering, University of Science and Technology of China, Hefei, 230026, China*

[2] *National Center for Protein Sciences Shanghai, Institute of Biochemistry and Cell Biology, Shanghai Institutes for Biological Sciences, Shanghai, 200031, China*

[3] *Centro de Investigaciones Opticas (CONICET La Plata-CIC), Cno. Centenario y 506, P.O. Box 3, 1897 Gonnet, La Plata, Pcia. de Buenos Aires, Argentina*

[4] *Physics Experiment Teaching Center, School of Physical Sciences, University of Science and Technology of China, Hefei, 230026, China*

Needle-like electromagnetic field has various advantages for the applications in high-resolution imaging, Raman spectroscopy, as well as long-distance optical transportation. The realization of such field often requires high numerical aperture (NA) objective lens and the transmission masks. We demonstrate an ultralong needle-like focus in the optical range produced with an ordinary lens. This is achieved by focusing a symmetric Airy beam (SAB) generated via binary spectral modulation with a digital micromirror device (DMD). Such amplitude modulation technique is able to shape traditional Airy beams, SABs, as well as the dynamic transition modes between the one-dimensional (1D) and two-dimensional (2D) symmetric Airy modes. The created 2D SAB was characterized through measurement of the propagating fields with one of the four main lobes blocked by an opaque mask. The 2D SAB was verified to exhibit self-healing property against propagation with the obstructed major lobe reconstructed after a certain distance. We further produced an elongated focal line by concentrating the SAB via lenses with different NA's, and achieved an ultralong longitudinal needle focus. The produced long needle focus will be applied in optical, chemical, and biological sciences.

## I. INTRODUCTION

An optical system is characterized by the point spread function (PSF), which measures the response of the optical instrument to an input spot. The PSF for an ordinary system exhibits an Airy-like disk with central lobe much stronger than the side lobe, and such distribution is employed to accurately localize the fluorescent molecule in super-resolution microscopies[1]. Sharper focus is absolutely required to resolve fluorescent molecules. In stimulated emission depletion microscopy (STED), a spot with narrow side lobe inhibits the fluorescence in the outer ring and enables super-resolution of the fluorescent molecules[2]. Various geometries of optical focus are realized through focus engineering. The optical needle b-

---

a) Electronic mail: yxren@ustc.edu.cn.

b) Electronic mail: lrd@ustc.edu.cn.

eam is such kind of focus featuring with the narrow electromagnetic filed distribution and the elongated longitudinal focal line[3, 4]. Such kind of needle beams is extremely valuable in various applications including the single-molecule fluorescence detection, particle transportation, and laser spectroscopy. However the generation of needle focus relies on the fine regulation of ring-shaped corrugated phase masks.

On the other hand, Berry and Balazs initially predicted an ideal nondiffracting Airy-like wave packet from the Schrödinger equation[5, 6]. The experimentally generated Airy-like beam was realized in optics by modulating the cubic phase in the spectrum space with liquid-crystal spatial light modulator (LC-SLM)[7, 8]. Such Airy beam exhibits Airy like pattern in the transverse plane and a ballistic trajectory for the central lobe during propagation[9]. The Airy beam possessing gradient force field acts as a snow-blower enabling the clearing of particles in a specific spatial region[10]. The self-accelerating Airy beam created with femtosecond intense pulses could induce a bent plasma channels[11] for conductive purpose. To satisfy different applications, variant forms of optical modes were experimentally created based on Airy functions, including the dual Airy beam[12], multiple Airy beam, and the abruptly autofocusing circular Airy beam (CAB)[13, 14]. Direct generation of the Airy beam from laser cavity could be realized by imprinting an aperiodic binary grating on the output coupling mirror[15]. The Airy modes also exist in the form of non-diffracting surface plasmon[16-18], self-accelerating acoustics[19], electron beams[20], as well as spatiotemporal non-spreading light bullets[21].

In the present work, an ultra-long needle focus was experimentally generated through the creation of a rectangularly symmetric Airy beam (SAB)[22, 23] with binary spectral amplitude modulation. The SAB firstly autofocuses and appears with a single central lobe, then the lobe splits into four Airy main lobes, and it looks like a combination of four Airy beams with outward acceleration[13, 14]. Different from the reported multiple Airy beams and abruptly autofocusing circular Airy beams, whose central lobes are closer to the coordinate center than the side lobe tail, the proposed SAB exhibits transversely rectangular symmetry with the side lobe tails surrounded by four major lobes. The creation of the SAB is implemented through an amplitude digital micromirror device (DMD)[24, 25]. The advantage of the DMD is its ability to presenting the diffractive patterns with a considerably fast speed[26]. Full control over the phase and the amplitude is achieved through binary Lee hologram[27]. With such amplitude SLM, we were able to demonstrate the gradual transition of the SAB from one-dimensional (1D) to two-dimensional (2D). We were able to create the ultralong needle focus via concentrating the SAB with lenses of various numerical apertures (NA's).

## II. THEORETICAL ANALYSIS

Airy wave packet was initially predicted By Berry and Balazs through deducing a special solution for the Schrödinger equation. The non-diffracting Airy packet with infinite energy could never be experimentally created. By truncating the wave packet, the finite-energy Airy beam was first experimentally realized in the form of optics. The subsequent studies demonstrated that such wave packet exists in various forms including the electrons, surface plasmons, and acoustics. In order to analyze the behaviors of Airy packet, we introduce the normalized paraxial equation of diffraction[7]:

$$i\frac{\partial u}{\partial \xi} + \frac{1}{2}\frac{\partial^2 u}{\partial s^2} = 0 \quad (1)$$

where $u$ denotes the electric field envelope, $s = x/x_0$ is a dimensionless transverse coordinate, $x_0$ represents an arbitrary transverse scale, $\xi = z/kx_0^2$ is a normalized distance in the direction of propagation, $k = 2\pi n/\lambda_0$ is the wavenumber in vacuum. Theoretical Airy beams are of infinite energy but not experimentally realizable. Experimental Airy beams are with finite power, and this is fulfilled by adding an exponentially decaying aperture function[7]. Such truncated Airy beam in the transverse plane is described as follows,

$$u(s, \xi = 0) = Ai(s)exp(as) \quad (2)$$

where $a$ is the decaying factor to truncate the tail of the infinite Airy beam. The corresponding Fourier spectrum in the spectral space is expressed as,

$$U_0(k) = exp(-ak^2)exp\left(\frac{i}{3}(k^3 - 3a^2k - ia^3)\right) \quad (3)$$

This implies that the Fourier spectrum of the finite Airy beam involves a Gaussian amplitude envelope and a cubic phase. Most of the incident beams are with a Gaussian transverse shape, and the truncating factor $a$ is considerably small, e.g., $a \ll 1$. The 1D Airy beam is experimentally obtained by modulating the incident Gaussian beam with a cubic phase spectrum, e.g., $\varphi_0(k) = exp\left(\frac{i}{3}k^3\right)$. The counterpart for 2D Airy beam is with the form of $\varphi_0(k_x, k_y) = exp\left(\frac{i}{3}(k_x^3 + k_y^3)\right)$. Therefore, the truncated Airy beam can be experimentally generated through modulating a Gaussian beam with a cubic phase[7, 9]. However, by replacing the spectral coordinates with the absolute value in the spectral cubic phase term, i.e., $\varphi(k_x, k_y) = exp\left(i\left(|k_x|^3 + |k_y|^3\right)/3\right)$, we can produce a 2D SAB with its intensity distribution exhibiting four major lobes situated on corners of a square[22]. The resultant beam is given by,

$$u(x,y,z) = \frac{1}{2\pi}\iint_{-\infty}^{+\infty} exp(-a(k_x^2 + k_y^2))exp\left(\frac{i}{3}\left(|k_x|^3 + |k_y|^3\right)\right)exp(iz\sqrt{k^2 - k_x^2 - k_y^2})exp\left(i(k_x x + k_y y)\right)dk_x dk_y \quad (4)$$

The creation of complex amplitude relies on the encoding of the field with a hologram. Such technique is usually based on off-axis holography and computer algorithms[27, 28]. The binary Lee method is an effective means to encode the desired complex field with binary amplitude hologram. We take the advantage of the spectral modulation of the cubic phase and the binary Lee method to produce the hologram for shaping the SAB. The continuous Lee hologram $t(k_x, k_y)$ for the desired phase $\varphi(k_x, k_y)$ is expressed as:

$$t(k_x, k_y) = \frac{1}{2}\{1 + cos[2\pi(k_x - k_y)v_0 + \varphi(k_x, k_y)]\} \tag{5}$$

where $(k_x, k_y)$ represents the spectrum coordinates on the surface of the DMD. A carrier with frequency $v_0$ is adopted to separate the light beams from different diffraction orders, and the SAB is produced in the first diffraction order[28, 29]. Such amplitude hologram provides continuous modulation of the incident beam. In order to accommodate the binary modualtion of DMD, a binary rounding technique is employed to optimize the continuous amplitude hologram. The binary version of the Lee hologram $h(k_x, k_y)$ is given by:

$$h(k_x, k_y) = \begin{cases} 1, & t(k_x, k_y) > 1/2 \\ 0, & otherwise \end{cases} \tag{6}$$

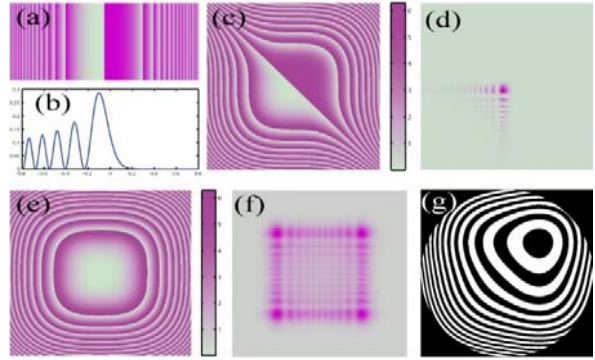

FIG.1. The cubic phase and beam profiles for 1D, 2D and SAB. (a) The 1D cubic phase structure produces 1D Airy beam (b). The 2D cubic phase (c) and the corresponding 2D Airy beam (d). Cubic phase (e) for SAB (f). The binary version of the Lee hologram for the creation of SAB.

The cubic phase mask for creating the 1D Airy beam and the intensity of theoretically predicted 1D Airy beam are respectively demonstrated in Fig.1 (a) and (b). 2D versions of the cubic phase mask as well as the theoretical intensity distributions are shown in Fig. 1(c) and (d) in sequence. By replacing the spatial coordinates with the absolute value, the cubic phase mask exhibits rectangular symmetry as shown in Fig. 1(e). Such symmetric cubic phase mask shapes the coherent collimated laser beam into SAB. The beam initially goes through a process of autofocus, and then its transversal profile spreads in the propagation space with a shape of four main lobes situated on corners of a square, which is theoretically shown in Fig. 1(f). By applying the Lee methods, the cubic phase for an SAB could be loaded on a carrier wave. Binary version of the computer-generated hologram (CGH) is demonstrated in Fig. 1(g). Since the produced hologram is square, we further pad the outside with 0's in order to fit the square pattern on the DMD chip.

## III. EXPERIMENTAL SETUP

The experimental setup for generating the SAB is shown in Fig.2. A single mode 633nm He-Ne laser (HJ-1, Nanjing Laser Instrument Company) with output power 1.5mW is utilized as the coherent light source. In order to fill the surface of the DMD, the coherent beam is expanded and collimated through a telescope composed of dual lenses and a pinhole spatial filter. The focal lengths of lenses L1 and L2 are 2.5cm and 20cm respectively. Lens L3 (24.3cm) acts as a Fourier lens that collects the modulated light to the spatial space at the back focal plane. A spatial filter allows the pass of only the first diffraction order of the modulated light. The produced intensity patterns are recorded by a CMOS camera (Weiyu Corporation, DS-CFM300-H, pixel resolution 2048×1536, pixel size 3.2μm) fixed on a guide rail. The beam profiles are saved on a local computer for further analysis.

The DMD (DLP7000, XGA, Texas Instrument) consists of a matrix of 1024×768 highly reflective and digitally switchable micromirrors[30]. Each micro-mirror with side length 14.46μm is connected with an underlying circuit board for further control through the computer. The micro-mirrors are situated on the parking/flat state when the power is off. By turning on the DMD, the micro-mirrors will rotate either +12° or -12° with respect to the main diagonal corresponding to the addressing electric signal '1' and '0'. These two kinds of angular positions correspond to a defined "on" or "off" state, which represents reflecting the incident beam to the designed optical path or sending the output beam to the light absorber, respectively. Therefore, each mirror of DMD individually serves as a light switch[31, 32], the DMD has numerous freedom to modulate the light field. With the amplitude DMD, several optical modes were successfully shaped, including the Laguerre-Gaussian[33], Bessel modes[34], flat-top modes[35] and Ince-Gaussian modes[26]. The shaped modes are of high fidelity and the DMD could facilitate the dynamic control.

The advantages of the DMD over LC-SLM are relatively higher fill factor and significantly fast switching rate. The fill factor for commercial LC-SLM is far less than 90%, as a comparison, the fill factor for the DMD is higher than 92%. The high fill factor makes better utilization of the incident light and suppresses the light shining on the dead area[26]. The fast refreshing rate of approximate 5.2 kHz enables dynamical control over the laser modes with a considerably fast speed. We were able to dynamically create the smooth transverse SAB modes. The hologram has to be numerically calculated, and we can easily acquire the desired optical profile at imaging plane. As a result, the DMD is more convenient to generate the beams with unique features as a spatial light modulator. The calculated hologram is directly loaded onto the DMD, and we eventually construct the SAB in the spatial Fourier plane of the collection lens.

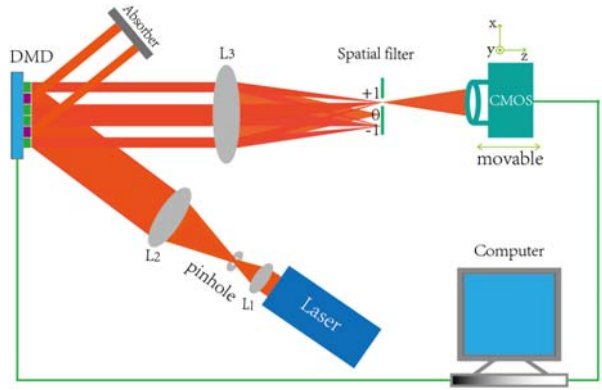

FIG.2. Schematic of experimental setup for creating the SAB. The output of the laser beam is filtered and expanded to slightly overfill the DMD front surface. The modulated light is collected by a lens L3 and the first diffraction is selected by a spatial filter. The CMOS camera mounted on a movable guide rail digitizes the transverse beam profiles at different positions on the propagation axis. A local computer both controls the DMD and records beam profiles from the CMOS camera.

## IV. RESULTS AND DISCUSSIONS

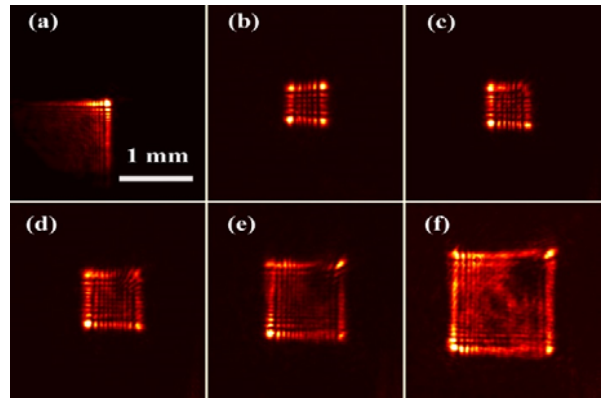

FIG.3. Intensity profiles of experimentally created (a) 2D Airy beam and (b) SAB. (c) The bottom-right major lobe of the SAB is obstructed by an opaque mask. The obstructed branch of the SAB will self-recover during propagation. Experimental observation of the intensity profiles at distances (d) $z=2$cm, (e) $z=4$cm, and (f) $z=6$cm, respectively. The scale bars for all the profiles are identical as shown in (a) for 1mm.

We were able to create the 1D and 2D Airy beam as well as the SAB. Fig. 3(a) shows the produced 2D Airy beam with the DMD displaying a pattern calculated through the binary spectral modulation. Fig. 3(b) is the experimentally created sectional beam profile for an SAB. A circular opaque mask was made by depositing an ink drop on the glass cover slide. Such opaque mask with diameter of approximately 300 microns was adopted to obstruct the bottom-right major lobe of the SAB. The produced SAB exhibits self-healing property[36] and is experimentally validated as shown in Fig. 3(c~f). The lobe immediately disappeared after the opaque mask. The disappeared major lobe was further reconstructed against propagation. The transverse intensity pattern for the recovered beam are shown in Fig. 3 for propagating distances of (d) 2cm, (e) 4cm, and (f) 6cm respectively. Direct comparison of the patterns shows that the blocked lobe gradually recovers during propagation distance. This is a clear manifestation that the produced SAB is self-healing.

The produced SAB could deform with different ellipticities. We introduce the directionally weighted 2D SAB with a parameter $\beta$ in the expression of spectral phase for 2D SAB to control the ellipticity. By replacing the spectrum coordinate $k_y$ with $\beta k_y$, i.e., $\varphi(k_x, k_y) = |k_x|^3 + |\beta k_y|^3$, we can construct a series of 2D directionally weighted elliptic SABs. The analytic expression of reconstructed elliptic 2D SAB mode will be,

$$u(x,y,z) = \frac{1}{2\pi} \iint_{-\infty}^{+\infty} exp\left(-a(k_x^2 + k_y^2)\right) exp\left(\frac{i}{3}\left(|k_x|^3 + |\beta k_y|^3\right)\right) exp\left(iz\sqrt{k^2 - k_x^2 - k_y^2}\right) exp\left(i(k_x x + k_y y)\right) dk_x dk_y$$

(7)

The parameter $\beta$ regulates the shape of the elliptic SAB. While $\beta = 0$, the elliptic SAB exhibits a homogeneous intensity in the y direction with an Airy-beam characteristics in the horizontal direction. While $\beta = 1$, the elliptic SAB is reduced to the conventional 2D SAB. We further demonstrated experimentally the dynamic transition among elliptic SAB modes as shown in the supplementary movie (Multimedia view)[URL will be inserted by AIP]. Fig. 4 shows intensity profiles of theoretically simulated 2D elliptic SAB with parameters (a) $\beta = 0$, (b) $\beta = 0.5$, (c) $\beta = 0.8$, (d) $\beta = 1$, respectively, and the corresponding experimental profiles are shown in Fig. (4), (e~h). These generated beam profiles are in a good agreement with the corresponding theoretical simulations.

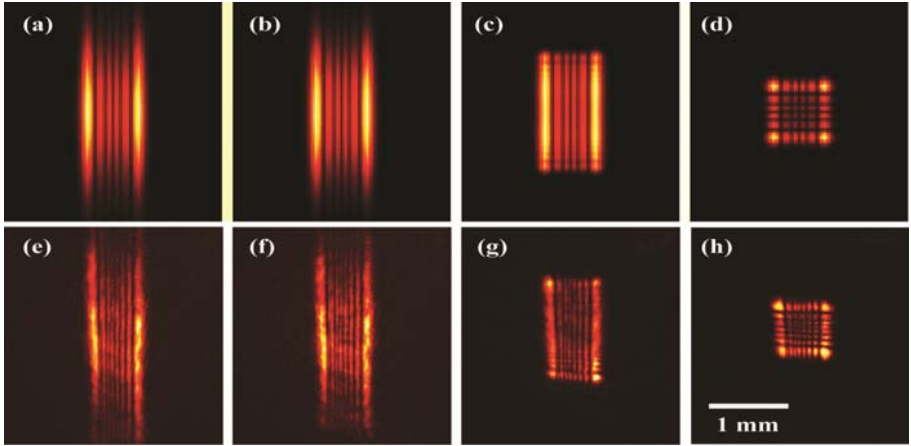

FIG.4. Snapshots for the transitional modes from 2D elliptic SAB to 2D SAB with various elliptical parameters. (a), (b), (c) and (d) are the theoretical simulations with control parameter $\beta = 0$, 0.5, 0.8, and 1, respectively; (e), (f), (g) and (h) are the corresponding experimental results. A supplemental media file[URL will be inserted by AIP] demonstrates the dynamical transition of SAB from (a) to (d).

The produced SAB is first collimated by an additional lens L4 (f=100mm, not shown in Fig. 2). We further employ the collimated SAB to form a sharp optical needle-like focus around the focal plane of the focusing lens L5. We found that the needle beam could be created with focusing lenses under various numerical apertures (NA's). The effective diameter for the modulated beam on the surface of focusing lens is approximately 4mm. Accordingly, the numerical apertures (NAs) of the lenses are 0.067, 0.040, and 0.025 respectively for the focusing lenses with focal length 80mm, 50mm, and 30mm. A CMOS camera situated on a movable guide rail records the sectional beam profiles with axial increment of 1mm (0.5mm for lens with 30mm focal length). All the measured sectional profiles, which are saved on the local computer for further analysis, constitute a data cubic. With this data cubic, we were able to reconstruct the longitudinal profiles of the light intensity around the focal spot.

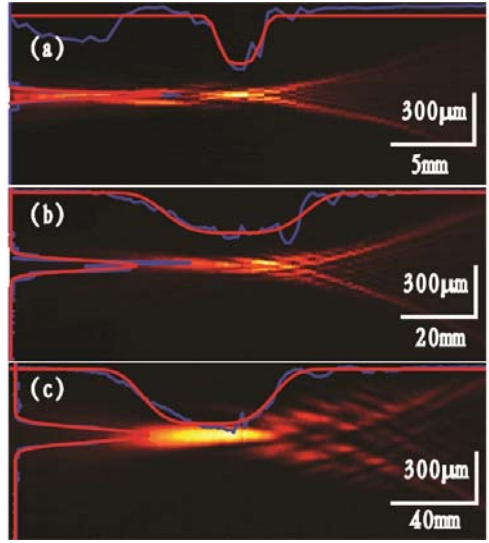

FIG.5. The experimentally created ultra-long needle focus with lens of different NA. The value of NA is changed via the focal length, while the effective entrance aperture for experiments with all lenses is 4mm. The focal lengths for all the lenses are (a) 30mm, (b) 50mm, and (c) 80mm respectively. The curves in each image demonstrate the transverse (left curves) and longitudinal (top curves) line profiles across the center of the needle. Blue curves represent the measurement, while the red curves denote the fitting with either Gaussian (transverse) or super-Gaussian beam mode (longitudinal) fitting.

We observed that the longitudinal needle of SAB can be enhanced by the lens with smaller NA. Fig.5 presents the x-z intensity profiles for lens with different NA. The longitudinal fields exhibit a needle-like focus for lenses with all of the three NAs. The needle focus is compared with previous reports on Hermite-Gaussian beams[37], the focus of which keeps the same transverse pattern but has reduced size (scale factor). The needle focus is unique for SABs. To quantitatively characterize the needle beam, the line profiles across the central spot are plotted in the respective intensity maps. We employ a super-Gaussian model $I = I_0 exp\left(-\left(\frac{z-z_0}{\omega}\right)^{2n}\right)$ to characterize the line profile. The full width at half maximum (FWHM) $FWHM = 2\omega \times \sqrt[2n]{\ln(2)}$ is employed to characterize the ultralong needle. The longitudinal line profiles are fitted with a 2-nd order super-Gaussian shape (n=2), while the transverse line profile is fitted with traditional Gaussian distribution (n=1). The blue curves represent the experimental measurements, while the red curves show the fitted curves with either Gaussian (transverse) or super-Gaussian (longitudinal) model. Accordingly, the FWHMs are 5.46mm, 34.84mm, 59.68mm along the longitudinal direction for lenses with focal lengths 30mm, 50mm, 80mm, and the corresponding FWHMs in the transverse direction are 42μm, 74μm, 96μm respectively. This indicates that smaller NA results in a longer needle but widens the transverse spot. The SAB mode possesses a paraxial watermark as the well-known HG beam. However, there are important differences between the SAB and the Hermite-Gaussian beams[37]. The SAB is a caustic beam with a "cusp structure" whose dynamics is governed by the three-ray superposition within the caustic region, limited by the caustic curve which divide the slit and shadow regions (annihilation of pairs of rays). The caustic classification arises from the phase symmetry rather than from the phase power, thus breaking the commonly accepted concept that fold and cusp caustics are related to the Airy and Pearcey functions, respectively[38]. The role played by the spectral phase power is to control the degree of caustic curvature. The even symmetry of the cubic phase is the generatrix of the cusp catastrophe structure. The major findings can be applied to any symmetric phase with order greater than 2, which is useful for increasing the performance in several applications. In spite of the qualitative similarities between HG's and SAB's in the far field intensity pattern, the former have no autofocusing properties which is a key characteristic of a cusp catastrophe structure due to the superposition of three rays in the caustic region.

## V. CONCLUSION

In conclusion, we have experimentally generated SAB through spectrum modulation with binary amplitude Lee hologram. The created SAB shares the self-healing ability, which was experimentally verified through measurement of the propagating field of the obstructed SAB. By introducing an elliptic parameter in the spectrum phase, dynamic control over the shape of SAB was achieved. Such dynamic control provides smooth transformation between directionally weighted SAB and 2D SABs. Furthermore, we have produced an elongated focal line by concentrating the SAB via lenses with different NA's, and achieved an ultralong longitudinal needle focus which is not obtainable with other higher-order spatial modes such as HG mode. The key difference is that the SAB possesses a cusp caustic structure while the HG mode does not. Such ultralong longitudinal focus could be applied in optical imaging Raman spectroscopy and micromanipulation.

**ACKNOWLEDGMENTS**

This work is sponsored by the National Natural Science Foundation of China (Grant No. 60974038) and Popularization of Science Foundation of Chinese Academy of Sciences (KP2015C10). PV acknowledges CNPq (Brazil) (Grant No. 311741/2014-2)